\documentstyle[emulateapj , epsfig,psfig]{article}
\def\ltsima{$\; \buildrel < \over \simlt \;$}
\def\simlt{\lower.5ex\hbox{\ltsima}}   

\def\simgt{\lower.5ex\hbox{>sima}}

\begin{document}

\title{Fragmentation of gravitationally unstable gaseous protoplanetary disks with radiative transfer}

\author{Lucio Mayer $^1$, Graeme Lufkin $^2$, Thomas Quinn $^3$, James Wadsley$^4$}

\affil{$^1$Department of Physics, Institut f\"ur Astronomie, ETH Z\"urich, Schaffmattrasse 10, 
CH-8093 Z\"urich, Switzerland, lucio@phys.ethz.ch} 
\affil{$^2$Department of Astronomy, University of Maryland, College Park, MD 
20742-2421, gwl@umd.edu}
\affil{$^3$Department of Astronomy, University of Washington, Seattle, WA 98195, USA, trq@astro.washington.edu}
\affil{$^4$Department of Physics \& Astronomy, McMaster University, 1280 Main St
. West
, Hamilton ON L8S 4M1 Canada,wadsley@physics.mcmaster.ca}

\begin{abstract}
We report on the results of the first 3D SPH simulation of massive, gravitationally unstable protoplanetary disks with radiative transfer. We adopt a flux-limited diffusion scheme justified by the high opacity of most of the disk. 
The optically thin surface of the disk cools as a blackbody. 
The disks grow slowly in mass 
starting from a Toomre-stable initial condition to the point at which they become marginally unstable. We find that gravitationally bound clumps with masses close to a Jupiter mass can arise. 
Fragmentation appears to be driven by vertical convective-like motions capable of transporting the heat 
from the disk midplane to its surface on a timescale of only about 40 years at 10 AU. 
A larger or smaller cooling efficiency of the disk at the optically thin surface can promote or stifle fragmentation by affecting the vertical temperature profile, which determines whether convection can happen or not, and by regulating the accretion flow from optically thin regions towards 
overdense regions. We also find that the chances of fragmentation increase for a higher
mean molecular weight, $\mu$, since compressional heating is reduced.
Around a star with mass $1 M_{\odot}$ only disks with $\mu \ge 2.4$, as expected for gas with a metallicity
comparable to solar or higher, fragment.
This suggests that disk instability, like core-accretion, should be more 
effective in forming gas giants at higher gas metallicities, consistent with the observed correlation between 
metallicity of the planet-hosting stars and frequency of planets.
\end{abstract}


\section{Introduction}

Long-lived clumps can form in a massive gravitational unstable protoplanetary disk if
the gas is isothermal (Mayer et al. 2002) or cools on a timescale comparable to
the local orbital time (Rice et al. 2003a,b, Mayer et al. 2004a, Rice et al. 2005, Mayer
et al. 2005, Mejia et al. 2005). Fragmentation under such favourable
conditions is a numerically 
robust result obtained with SPH codes as well as grid-based codes with both static and adaptive meshes
(Durisen et al. 2005, Mayer et al., in prep.). The main open question is 
whether the required short cooling times can be obtained in the disks. 
Shock heating along spiral arms is intense during the phase of strong gravitational instability (Pickett et al. 2000, 2003; Mayer et al. 2004b) and will tend to erase clumps exactly when they have the best chance to collapse.
Boss (2002a,2002b,2004) has used a grid-based code which solves radiation transport
with the  diffusion approximation. He found that the disk cools rapidly through 
convection instead of radiation and forms long lasting clumps.
Cai et al. (2006) adopt flux-limited diffusion in their cylindrical grid code. 
They find long cooling times, no evidence of convection and no fragmentation.
Their method is very different from that of Boss (2004). While Boss 
does not use a flux-limiter but embeds the disk in a thermal bath at a constant 
temperature in the range 40-50 K,  Cai et al add an Eddington-like atmosphere 
on top of the optically thick layer, explicitly matching the fluxes at the boundary. 
Nelson et al. (2000) also did not find fragmentation in a 2D SPH simulation to which
a plane-parallel atmosphere radiating as a blackbody was added assuming that rapid vertical cooling was
achieved through convection. Here we present the first 3D SPH simulations of protoplanetary disks 
with flux-limited diffusion. 

\section{The simulations}

The disk models are set up as described in Mayer et al. (2004). Initially
disks extend from 4 to 20 AU, have a power-law surface density profile and 
have a minimum temperature of 40 K at the outermost radius.
The simulations employ $10^6$ gas particles with a gravitational 
softening $0.06$ AU, while the central star is represent by a particle
with mass $1 M_{\odot}$ and softening  $0.2$ AU.
Disks are grown in mass from a stable state with a high Toomre parameter (Tommre 
1964), $Q_{min} > 4$. 
 Mass growth is stopped either when fragmentation begins or when the disk reaches a mass equal to $0.2 M_{\odot}$.
Thanks to the very high resolution we always fulfill the criterion of Bate \& Burkert (1997) to avoid
spurious fragmentation. 
The simulations were run with the parallel SPH+N-Body code GASOLINE (Wadsley, 
Stadel \& Quinn 2004). We solve the energy
equation in asymmetric form accounting for irreversible shock heating via the standard Monaghan artificial
viscosity term (we include the Balsara correction term to reduce unwanted shear viscosity).

The radiation transport is
implemented using the diffusion approximation and the flux-limiter of Bodenheimer et al. (1990). A
similar method has been used by Whitehouse \& Bate (2006) to study the collapse of molecular cloud cores.
The flux limiter appears as a coefficient  of a diffusive term in the energy equation. 
Following Cleary \& Monaghan (1999),
the energy equation reads 
\begin{equation}
\label{eqn:udot_sph}
  \dot{U}_a = \sum_b \frac{4 m_b}{\rho_a \rho_b} \frac{k_a k_b}{k_a + k_b} \left( T_a - T_b \right) \frac{\mathbf{r}_{ab} \cdot \nabla W}{|\mathbf{r}_{ab}|^2}
  \end{equation}
where the summation is over neighboring particles, $W$ is the smoothing kernel, $\mathbf{r}_{ab}$ the vector from the position of particle $a$ to particle $b$, and $k_a$ takes the form of a thermal conductivity term 
\begin{equation}
  \label{eqn:k}
  k_a = \frac{16 \sigma}{\rho_a \kappa_a} \lambda_a T_a^3 .
  \end{equation}

where $\kappa_a$ is the opacity, $\sigma$ is the Stefan-Boltzmann constant and
$\lambda_a$ is the flux limiter.
This is a stable method that relaxes the noise of second derivatives.

\medskip
\epsfxsize=8truecm
\epsfbox{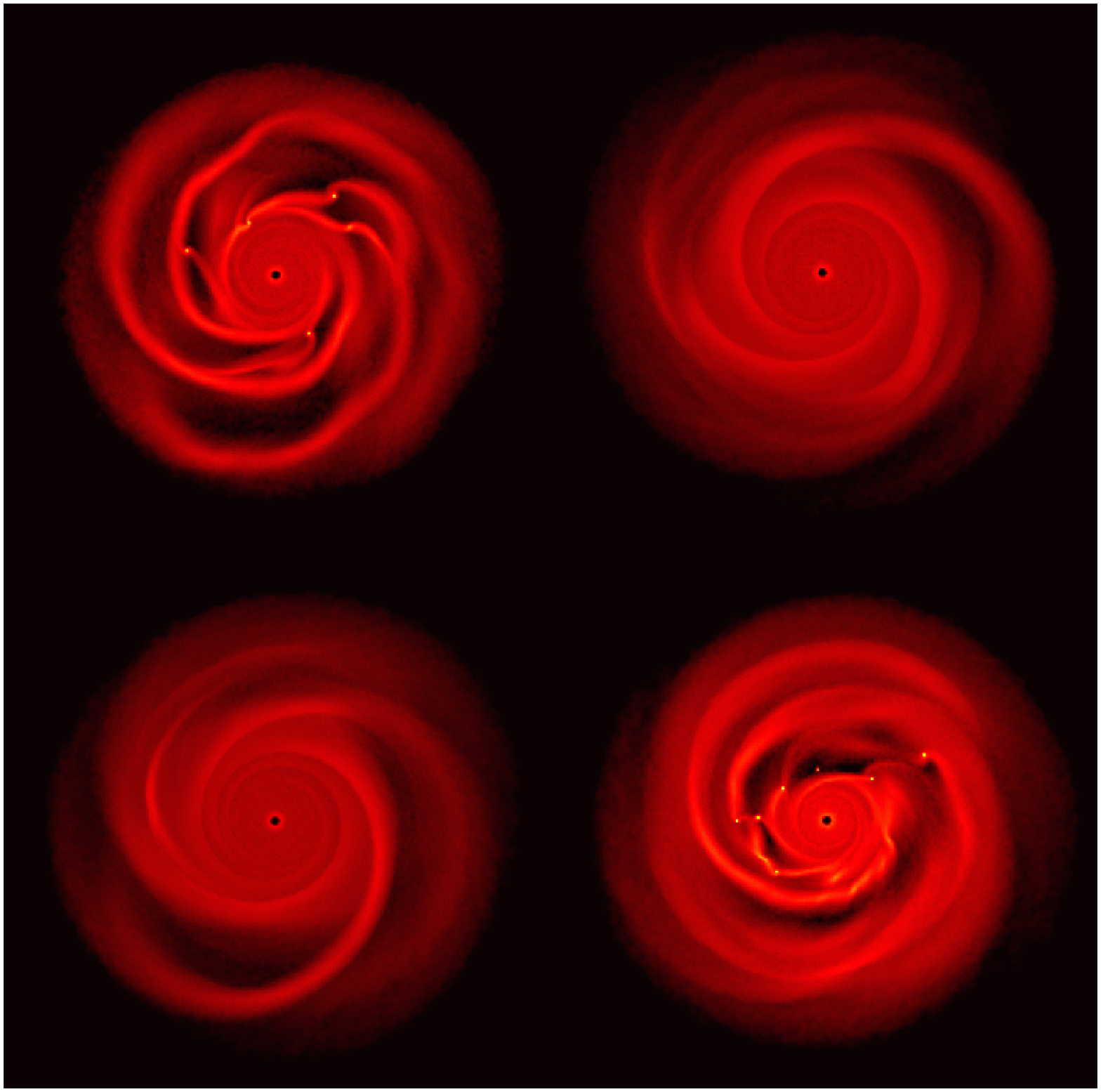}
\figcaption[mbar1.ps]{\label{fig:asymptotic}
\small{Color coded logarithmic density plots of the disk seen face-on after 1300
years of evolution. From top left to bottom right we show the runs with (a) $\mu=2.4$, EDA (edge detection angle)
=30 degrees, (b) $\mu=3$, EDA=60 degrees, (c) $\mu=2.4$, EDA=40 degrees,
(d)$\mu=2.7$, EDA=40 degrees. The maximum densities shown, at the clumps position, reach
values higher than $10^{-9} g/cm^3$.}}
\medskip

To implement the above expression in {\sc Gasoline}, we add the calculation of the temperature gradient to the smoothing operation already in place to calculate the force due to the pressure gradient.
The (Rosseland mean) opacity of each particle is interpolated from a table of $\kappa(T, p)$ (D'Alessio et al. 1997). 

The next step is to allow particles on the 
boundary of the disk to radiate their energy away to infinity.
To find particles that are ``on the edge'' of the disk, we examine the directions to all of the neighbors used in smoothing sums. If a particle has no neighbors within a certain fraction of a solid angle 
from a preferred direction (edge detection angle, hereafter EDA) it is considered an edge particle.
From the geometry of a disk, the preferred directions (treated independently) are out of the plane of the disk (both up and down) and radially outward.
We let edge particles radiate as blackbodies, adding a term to the energy equation of each 
particle,
\begin{equation}
\label{eqn:udot_black_body}
\dot{U}_a = f_a S \sigma T_a^4 / m_a .
\end{equation}
where $S=4\pi h_a^2$ is the surface through which the particle radiates and $h_a$ is the smoothing length of the particle.
The ``edgeness factor'' $f_a$ represents the fraction of their surface area over which a particle radiates. It is usually zero, and takes value $1/2$ 
for particles on one of the up, down, or out boundaries, $1$ for those
on the edge in both the up and down directions, and $3/4$ for those on the 
edge in the out and either up or down direction.

The EDA used in the check for ``edgeness'' is formally a free parameter. We only consider
angles in the range 30-60 degrees (measured as linear angle in a cone 
centered on a given particle)
since with these the particles identified as ``edge'' do lie in regions 
where the optical depth is $\tau < 1$.
Different choices for the EDA will effectively yield a different radiative efficiency since both the radiating surface area
and the edgeness factor for each particle will change. 
The smallest possible size of the radiating surface area
is that of the geometric surface area of the disk, which
corresponds to an EDA of 50 degrees when the disk enters the phase of gravitational
instability. 
Within the range of angles considered the radiative surface area changes by about 
a factor of 2.

The term in equation 3 has a timescale that can be several orders of magnitude 
smaller than the others.
To handle this, we apply a special Burlisch-Stoer stiff integrator to this term over 
a single timestep in the outer leapfrog integrator.
We hold the other terms of the energy equation constant over this short integration.

A near solar opacity is used in most of the simulations.
The (mean) molecular weight of the gas is by default equal to the solar metallicity 
value ($\mu=2.4$) but is varied between $\mu =2$ (pure molecular hydrogen) and $\mu=3$ (highly metal rich).  
We have run about 15 simulations with varying molecular weight and opacity.

\section{Cooling, heating and fragmentation}

The disk grows uniformly in mass at the rate of $\sim 10^{-4} M_{\odot}/$yr, approaching $0.1 M_{\odot}$ 
after about $10^3$ years. 
When its mass grows above $0.05 M_{\odot}$ the Toomre parameter $Q$ drops below 2 in the outer part of
the disk and strong spiral patterns begin to appear. 
The shocks occurring 
along the spiral arms limit the growth of their amplitude as the increasing
pressure counteracts self-gravity. Yet, fragmentation occurs
in some of the simulations once the disk mass is in the range $0.12-0.15 M_{\odot}$ (Figure 1).
Whether the disk fragments or not depends on  the details of thermodynamics in the disks, 
in particular the molecular weight of the gas 
and the extent of the radiative layer (see Figure 1). 

We find that a molecular weight
equal to  $2.4$  or larger
is in general necessary for fragmentation to occur (Figure 1). Larger molecular weights
effectively soften the pressure gradients across the spiral arms and allow fragmentation to happen for a smaller radiating surface area (Figure 1).
This sensitivity is reminiscent of the sensitivity on the adiabatic index $\gamma$  found by 
Mayer et al. (2004b) and Rice et al. (2005)(here we fix $\gamma=7/5$).
Conversely we varied the disk opacity by a factor of 50 
and found almost no difference in the outcome, confirming the results of Boss (2002a). 
Indeed when cooling at the surface is switched off the temperature in the midplane is the same as in 
a control run performed with an adiabatic equation of state. This is not surprising. The 
timescale over which thermal energy is transported from the midplane
to the surface, which is of order 
$10^4$ years (Boss 2003), is the product of the radiative diffusion time, which depends
on opacity,  times the ratio between gas pressure and radiation pressure.
The latter factor
is orders of magnitude higher than the radiative diffusion time, is 
independent of opacity,
and is what determines the radiative cooling time of most of the disk. 
This confirms earlier findings by Pickett et al. (2000, 2003) on the crucial
role of compressional/shock heating in disk instability.

However, the cooling time
has to be $\sim 100$ times shorter, namely of order of the orbital time, for fragmentation
to happen (Rice et al. 2003a; Mayer et al. 2004a). Such short radiative cooling 
times are only possible at the optically thin edge in our disks.
Boss (2002b, 2004) finds that the disk can cool on such short timescales
thanks to convection.
The optical depths in our disks are very large, $\tau > 10$, near the disk
midplane. Convection is expected (Ruden \& Pollack 1991) in such conditions.
The following sequence of events is observed in the simulations and translates
into the temperature evolution of overdensities  shown in Figure 2;
first, density maxima develop along an optically thick spiral arm in the
disk midplane; second, the gas in
the spiral arm is shock heated to temperatures above 150 K; third, if 
the disk atmosphere is cold enough while the midplane is shock heated
vertical upwellings and downwellings begin  which cool the 
clumps and allow them to become gravitationally bound.
Such vertical motions have typical speeds of order  ~0.1-0.4 km/s (the typical orbital speed
at the same radii is $\sim 10$ km/s). They involve 
most of the gas particles in a locally overdense region of the disk and 
occur when the vertical entropy gradient is superadiabatic. 
These motions where not 
observed in previous calculations with fixed equations of state in which steep vertical temperature gradients were not produced (Mayer et al. 2004b).
The pressure scale height, whose
scale should be comparable to the mixing length in the convective motions, is from 10 to 50 times 
bigger than the SPH smoothing length, hence convection should be well resolved.

The measured vertical speeds
are high enough to redistribute thermal energy on a timescale of about 30-50 years, i.e. from
comparable to slightly longer than the orbital time. With such short cooling timescales fragmentation
is expected (Mayer et al. 2004a; Rice et al. 2003a). It is difficult to measure the net
transport of energy due to convection because turbulent motions can arise
also as a result of shocks (Cai et al. 2006) and simultaneously vertical
motions are also produced by the accretion flow towards the overdensities
at the midplane.

\medskip
{\centering
\epsfxsize=5.5truecm
\epsfbox{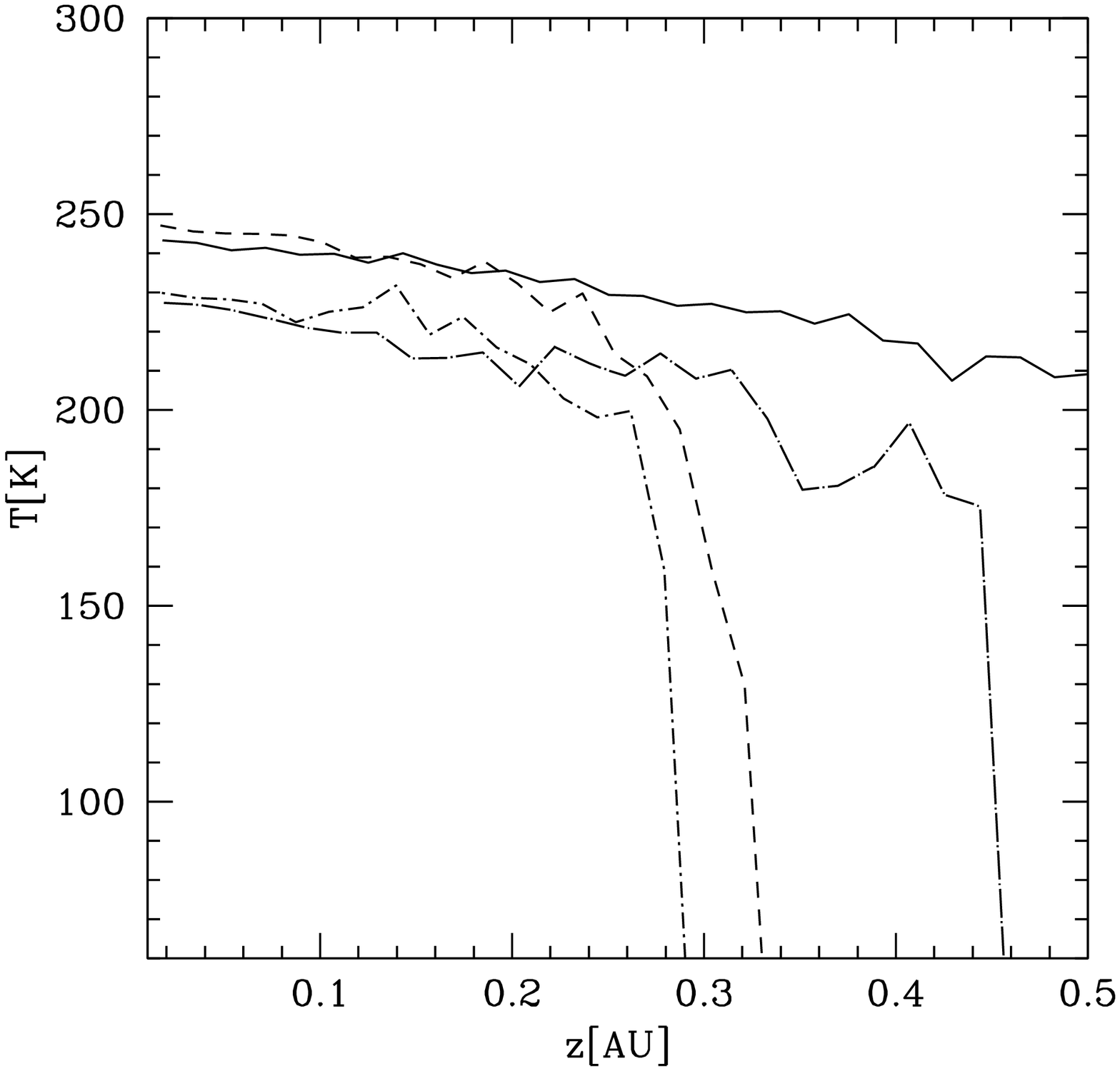}
\epsfxsize=5.5truecm
\epsfbox{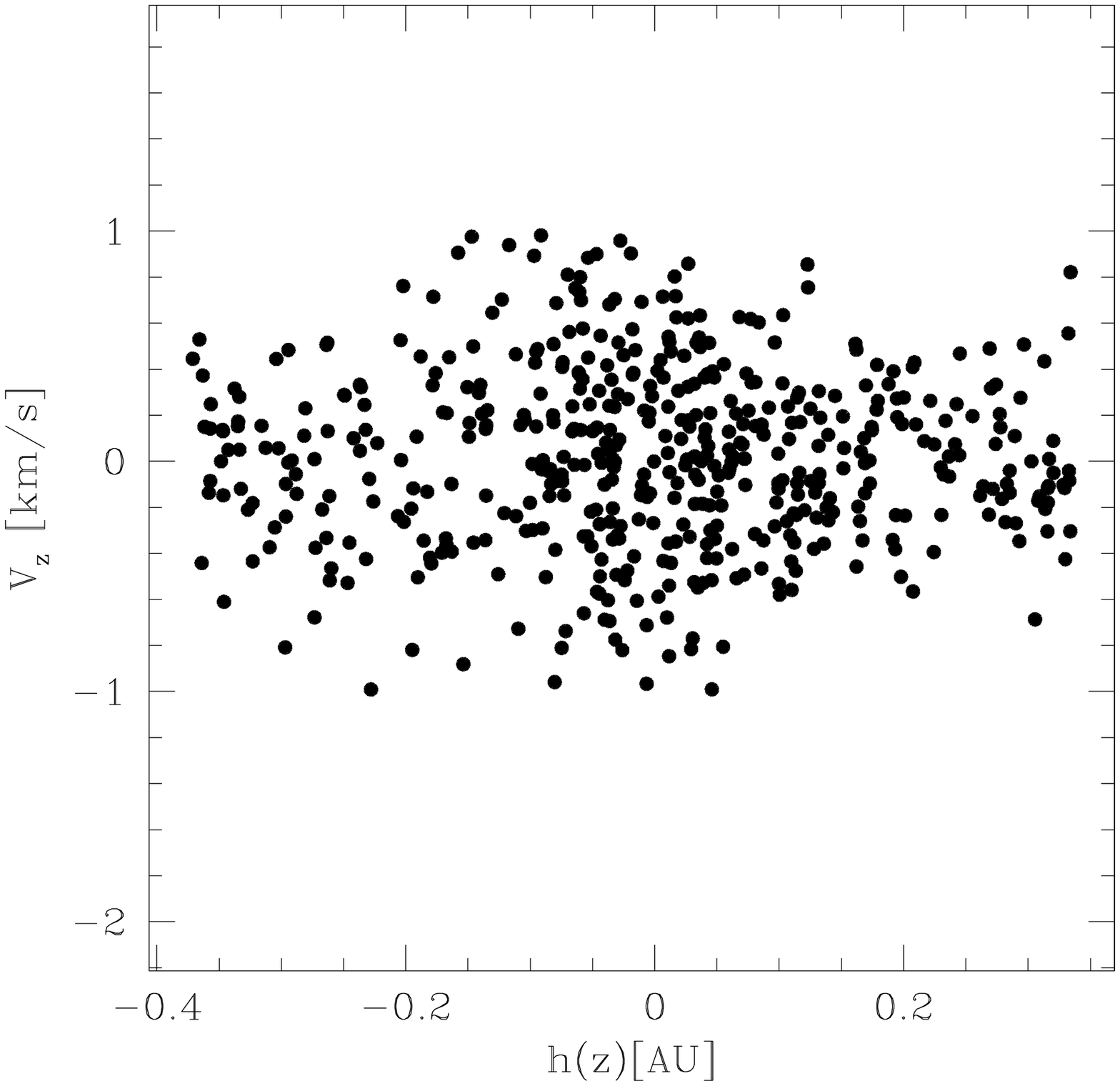}
\figcaption[mbar1.ps]{\label{fig:asymptotic}
\small{Top: Evolution of the vertical temperature profile of an overdense region that collapses and becomes
a gravitationally bound clump in the run with $\mu=2.7$ and EDA = 40 degrees (same clump
as in Figure 3). The profiles are taken at t$=1220$ years (solid line), t$=1221$years (dashed line), t$=1222$ years (short dot-dashed line) 
and t$=1223$ years (long dot-dashed line), and are azimuthally
averaged. As the profile becomes steeper the gas becomes convectively unstable. It then relaxes to a shallower, more stable
profile, with an average temperature lower than at the start (dot-dashed line). 
Bottom: Velocity component perpendicular to the disk midplane as a function of the distance from the center of mass of the clump 
at t$=1221$ years.}}}
\medskip

In addition to affecting convection by changing
the vertical temperature profile, the radiative efficiency at the cold 
optically thin layer has a second influence on the final outcome of the disk 
instability.
The dense regions along the spiral shocks can engulf material from colder regions outside
the shock front (Figure 3),
and this accretion flow is stronger for colder material. 
These regions
are optically thin down to a vertical scale height half of that of the neighboring overdense spiral arms and are naturally produced as the arms unwind. 
They provide a reservoir of cold gas that overdensities can accrete, promoting their growth.
This is an example of the complex coupling between the dynamics
and thermodynamics seen in these simulations. Nevertheless, we notice that
the typical midplane temperatures ($> 200$K) and midplane surface densities 
($\sim 2000$ g/cm$^{2}$) of our disks where clumps first appear 
(at about $15$ AU) are admitted as fragmenting solutions by the analytical
model of Rafikov (2005), given a Toomre parameter $Q=1.4$ and
a cooling time $t_{cool}=1.5 T_{orb}$ as the criteria for fragmentation
(see Mayer et al. (2004a).

Once formed, gravitationally bound clumps
have masses that range range from one to a few Jupiter masses, are differentially
rotating and have densities a million times higher than the background density.
They reach $\tau \sim 100$ at the scale of the softening and temperatures  $T > 300$ K.
In general they resemble clumps formed in simulations with an adiabatic equation of state 
above a fixed density threshold (Mayer et al. 2004a,b).

\section{Discussion}

We have shown that disks can fragment even when radiative transfer is included. 
However, fragmentation requires a disk somewhat more
massive than in previous works relying on simple equations of states or parameterizations of the
cooling time., $M > 0.12 M_{\odot}$ as opposed to
$\sim 0.1 M_{\odot}$ (Mayer et al. 2004a,b). It is strongly dependent on
the efficiency of cooling at the optically thin boundary and  on the molecular weight.
The newly discovered dependence  on molecular weight is intriguing 
since it establishes for the first time a direct link between the disk instability mechanism and the metallicity 
of the gas.  Such a link is straightforward in the competing model, core-accretion (Pollack et al. 1996).
Hence the formation of giant planets by disk instability should be favoured around more metal rich
stars, in line with the observed correlation (Fischer \& Valenti 2005).
A more metal rich disk will also have a higher cooling rate in the optically thin region (Cai et al. 2006),
which also appears to promote gravitational instability,
whilst the associated increase of opacity in the optically thick regions will not affect
the outcome based on our results.

\medskip
\medskip
{\centering
\epsfxsize=8.5truecm
\epsfbox{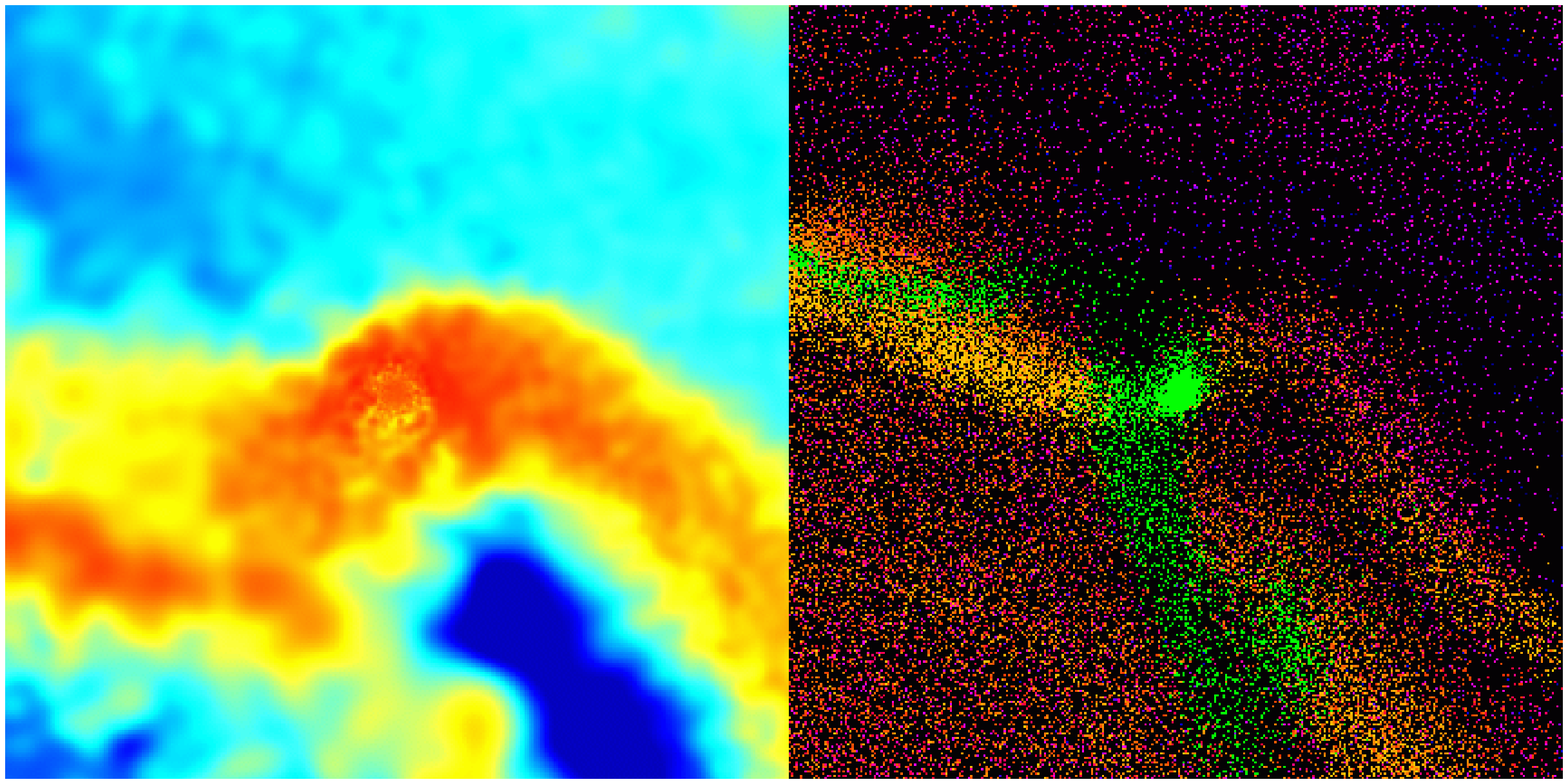}
\figcaption[mbar1.ps]{\label{fig:asymptotic}
\small{Color coded logarithmic temperature plots of a region of size about 4.5 AU centered on one of the
gravitationally bound clumps in the run with $\mu=2.7$ and EDA=40 degrees. Left: a smoothed
temperature map is shown (the temperature range goes from 10 K (dark blue) to 250 K (dark red) at time
$t=1248$ years showing the cold, low density regions adjacent to the hot gas in the spiral shock and
clump. Right: temperature map of the gas particles surrounding the same clump about half an orbit earlier
with the particles that will accrete by $t=1248$ years marked in green. The brighter the color the higher
the temperature. Most of the accreted mass appears to come from the colder regions outside the spiral shock.}}}
\medskip

The radiation physics in the simulations presented here is still simplified. In particular,
the disk edge could be heated as a result of irradiation from the central star and/or neighboring bright stars, 
erasing the steep temperature gradient necessary to drive convection.
On the other end, when an overdensity forms dust might accumulate at its location owing to the action of gas drag
(Haghighipour \& Boss,. 2003, Durisen et al. 2005), thus increasing locally the 
molecular weight and enhancing the chances of fragmentation.
In the latter case the clumps would be automatically enriched in heavy elements
relative to the surrounding disk, as seen in the gas giants of the Solar System (Saumon \& Guillot 2004) 
and predicted by recent models of core accretion (Alibert et al. 2005).

There are situations in which fragmentation may be achieved more easily than 
suggested here. Indeed a new set of simulations being
completed as we write indicates that if the mass of the 
central star is lowered to $M=0.5 M_{\odot}$, the disk fragments for masses $< 0.1 
M_{\odot}$ for a molecular weight as small as $\mu = 2$.
In this case a lower shear allows a higher pressure 
front to be tolerated from growing overdensities. Therefore with disk instability formation of
giant planets around M dwarfs is likely (see also Boss 2006). We also find that if the disk does not
grow slowly in mass but starts with $Q_{min}=1.4$, as it might happen in response to a sudden
external perturbation, fragmentation occurs with $\mu=2$ and edge detection angles of up to 
60 degrees. On the other end, previous work  shows that long lasting, time-dependent perturbations
such as those induced by the presence of a binary companion might quench the
fragmentation by increasing 
the heating due to spiral shocks (Mayer et al. 2005).

In summary, it seems that an ultimate answer requires that
all the details of the radiation physics in the disks and in their surrounding environment
are taken into account together with the details of the disk formation process.

\bigskip

L.M. thanks Richard Durisen, Willy Benz, Annie Mejia, Willy Kley, Yann Alibert and Laure Fauchet for fruitful discussions.
TQ acknowledges support from the Nasa Astrobiology Institute.
Simulations were performed on LeMieux at the Pittsburgh Supercomputing Center and on Zbox at the University of Zurich.

\end{document}